\documentstyle[12pt]{article}
\begin{document}
\title{Calculating Masses of Pentaquarks Composed of Baryons and Mesons}
\author{  \thanks  \ M. Monemzadeh\thanks{monem@kashanu.ac.ir},
N. Tazimi \thanks {nt$_{-}$physics@yahoo.com}, 
 \ Sh. Babaghodrat\thanks{shbabaghodrat@yahoo.com}, \\
\it\small{{Department of Physics, University of Kashan, Ghotb Ravandi Boulevard, Kashan, Iran }}}
\date{}
\maketitle
\begin{abstract}
In this paper, we consider an exotic baryon (pentaquark) as a bound state of  two-body systems composed of a baryon (nucleon) and a meson.We used a  baryon-meson picture to reduce a complicated five-body problem to a simple two-body problems. The homogeneous Lippmann-Schwinger integral equation is solved in configuration space by using  one-pion exchange  potential. We calculate the  masses of pentaquarks  $\theta_{c} (uudd\bar{c})$, $\theta_{b} (uudd\bar{b})$
 
\end{abstract}
 \textbf{Keywords:} Exotic baryon, pentaquark, Lippmann-Schwinger equation,one-pion exchange potential  \\

PACS Numbers:12.39.Jh, 12.39.Pn, 14.20.Pt, 21.10.Dr, 21.45.+v

\newpage 
\section{Introduction}
There are two types of hadrons, baryons and mesons. Baryons are equivalent to the bound states of three quarks, and mesons are known to be the bound states of a quark and an antiquark. However, QCD describes mesons and baryons even with a more intricate structure. There are anomalous  mesons such as   $ggg$, $q\bar{q}g$, $q\bar{q}q\bar{q}$, etc. as well as  exotic baryons like $qqqg$, $qqqq\bar{q}$, etc. Pentaquarks are baryons with at least four quarks and one antiquark. In exotic pentaquarks, the antiquark has a flavor different from the other four  quarks. \\

Exotic hadrons containing at least three valence quarks are being studied fairly extensively in modern physics. Although there are hundreds of ordinary hadrons, exotic ones haven't been found stable yet. However, QCD does not reject their existence.
Pentaquark $ \theta^{+} $, studied in photo production experiments \cite{1,2}, is a prototype of exotic hadrons in light and strong quark sector. Theoretically, hadronic reactions contribute to $\theta^{+} $ production more vividly than other types of reactions.\\

The quark model is commonly used to describe hadrons. In this model, mesons are described as $q\bar{q}$ and baryons as three-quark composite particles. In a more microscopic view, QCD usually serves to describe the strong interaction.
According to Lipkin \cite{3,4}  and Gignoux et al. \cite{5}, among pentaquarks, the five-quark anticharmed baryons of the $P^{0}=[uud\bar{c}s] $ and $\bar{p}=[udd\bar{c}s] $ or similar anti-beauty baryons are the  most bound. \\
A lot of experimental evidence on the existence of exotic hadrons has been found since 2003. Exotic hadrons' quantum numbers cannot be  justified based on two- and three-quark bound states. Pentaquarks of $qqqq\bar{q}$ form are examples of exotic baryon states. Conjugation quantity of $C$ charge is not an accurate quantum number for baryons, and all combinations of total spin $J$ and parity $P$ can exist. However, an exotic baryon combination can be readily identified by its electric charge $Q$ and its strangeness $S$.
  Some evidence has been reported during the last few years. For example, the pentaquark  $ \theta^{++} $ was proven to exist in Hermes experiment in Humburge, Germany \cite{6,7}.\\ For exotic baryons, we consider the following:\\
$ \theta^{+}$: The existence of  this exotic baryon was predicted in chiral solution model  \cite{8}. 
It has an  $S=+1$,  $J^{P}  =1/2^{+}$   and $I=0$. It is a narrow light-mass particle of 1540 MeV. These attributes initially made $ \theta^{+}$  a subject of  experimental observation by LEPS  \cite{9} .The most  suitable hadronic decay mode to identiy it
 is $ \theta^{+}$  $ \longrightarrow K^{0}p$.
 
$\theta_{c}$ and $\theta_{b}$: The existence of the bound exotic hadron $\theta_{c}$ was predicted through bound  Skyrmion approach. This particle has a mass of 2650 MeV, quantum numbers  $J^{P}  =1/2^{+}$  and  $I=0$. An experiment \cite{10} showed a positive signal at a mass of about 3.1 GeV, but it wasn't confirmed  later \cite{11}. In strongly bound states, the decay mode  $K^{+}\pi^{-}\pi^{-}p$ is easy to identify.
Likewise, the mass of $ \theta_{b}$ with the same quantum  numbers $J^{P}=1/2^{+}$   and  $I=0$ was predicted to be 5207 MeV. The possible weak decay made $K^{+}\pi^{-}\pi^{-}+\pi^{+}+p$.\\

$ \theta_{cs}$: It is the five-quark  state with $J^{P}  =1/2^{-}$ and  $I=0$. In a quark model which includes color-spin interaction, it can be bound and despite its strong decay, it becomes stable \cite{12}. The mass dependent on the model parameters is predicted to be  2420 MeV. This state was traced in the Fermilab E791 experiment via  $\phi \pi p $ mode \cite{13} and  $K^{*0}K^{-} p$   mode \cite{14}.\\      

  Lipmann-Schwinger Equation for two-body bound states is solved in sec 2.  In sec 3,  the procedure of the study is given and pentaquark masses are calculated.\\
 
\section{Lippmann-Schwinger equation for two-body bound states}
 In this part, the binding energy  of the entire system (pentaquark) is calculated by numerical solution of homogeneous Lippmann-Schwinger  equation for each subsystem of  bound meson and baryon. Schrodinger equation for a two-body bound state with the potential $V$ runs as the following integral equation \cite{16,17}:

\begin{equation} 
\psi_{b}(r)=-\frac{m}{4\pi} {\int_{0}^{\infty}}{ dr^\prime} {r^\prime}^{2}{\int_{-1}^{1}} dx^\prime {\int_0}^{2\pi} d\phi ^\prime \frac{exp(-\sqrt{m\vert E_{b}\vert} \vert{r-r^\prime}\vert)}{\vert{r-r^\prime}\vert} V(r^\prime){\psi_{b}(r^\prime)} 
\end{equation}
where $ E_{b} $ stands for the binding energy of the two-body bound system (meson+baryon). When the interaction potential is considered as $ r $ function, (\ref{1}) is of the following eigenvalue form:
\begin{equation}
K(E_{b})\vert\psi_{b}>=\lambda(E_{b})\vert\psi_{b}>
\label{2}
\end{equation}

$\lambda=1$  is the highest positive eigenvalue. The eigenvalue equation is solved through direct method  \cite{16,17,18,19} using a regularization method \cite{20,21}.  To discretize the integrals, Gauss-Legendre method \cite{22} is employed. 200 grid points is considered for $  r$ and $ r\prime $, whereas 100 mesh points is quite enough for $x\prime$.  The kernel of integral equation is diagonized to find $\lambda=1$ in the eigenvalue spectrum. The energy corresponding to $\lambda=1$ will be the system's binding energy.

\section{ Numerical Results and Discsussion} 
We consider a pentaquark as  a bound state of a two-particle system formed by a baryon and a meson. The hadronic molecular structure consists of a baryon and a meson is shown in Figure 1.

\begin{figure}[h]
\includegraphics{penta.jpg}
\caption{Pentaquark}
\vspace{6cm}
\label{1}
\end{figure}
Interaction potential has an essential role in solving the eigenvalue equation (5). Different potentials have been introduced for meson-baryon interaction. Yukawa potential   (screened coulomb potential) is proposed as one of the appropriate ones \cite{15,23}.
  This study deals with exotic baryon states created by a meson and a nucleon. The $ \pi $ exchange potential is among the most prominent meson exchange forces. $ \pi $ is the lightest hadron that can be exchanged between a meson and a nucleon. Therefore, we consider only $ \pi $ exchange, and $ \rho $ and $ \omega $ meson exchange will be elaborated on in our subsequent works. One-pion exchange potential (OPEP)  is of the following form[15]:\\ 
  
\begin{eqnarray}
 \quad  \quad  \quad \quad \quad V_{\pi}(r)=   \left \lbrace 
\begin{array}{c c } 
 (\vec{I}_{N}.\vec{I}_{H}) ( 2S_{12} V_{T}(r)+4\vec{S}_{N}.\vec{S}_{l}) V_{c}(r) \quad, r>r _{0}  \quad  \quad  \quad \quad   \quad  \quad \quad  \\
V_{0} \quad \quad \quad   \quad \quad \quad \quad \quad \quad\quad \quad  \quad  \quad  \quad \quad  r < r _{0} \quad  \quad   \quad \quad  \quad  \quad \quad 
 \end{array}
\right. \end{eqnarray} 
where for $ r> r _{0} $    :  \\
\begin{eqnarray}
V_{\pi}(r) &=&(\vec{I}_{N}.\vec{I}_{H}) ( 2S_{12} V_{T}(r)+ 4\vec{S}_{N}.\vec{S}_{l})V_{c}(r)
\label{11} \nonumber \\
& =&(I^{2}-I_{N}^{2}-I_{H}^{2})( 2S_{12} V_{T}(r)+ K^{2}-S_{N}^{2}-S_{l}^{2})V_{c}(r)
\label{11}
\end{eqnarray}
where :
\begin{equation} 
V_{c}(r)=\frac{g_{H}g_{A}}{2\pi f_{\pi}^{2}}(m_{\pi}^{2}) \frac{e^{-m_{\pi}r}}{3r}
\label{11}
\end {equation}

\begin{equation}
V_{T}(r)=\frac{g_{H}g_{A}}{2\pi f_{\pi}^{2}}(m_{\pi}^{2}) \frac{e^{-mr}}{6r}(\frac{3}{r^{2}}+\frac{3}{r}+1)
\label{11}
\end {equation}

\begin{equation}
V_{0}= -62.79 ( MeV)  or -276 ( MeV)
\end {equation}

where $g _{A}$ , $ f_{\pi}$, $m_{\pi}$, and $g _{H}$ are axial coupling constant,  pion decay constant, pion mass, and  heavy-meson coupling constant, respectively. $ I $ is the total isospin of meson-nucleon system and  
\begin{equation}
S_{12}\equiv 4[ 3 (\overrightarrow{S}_{N}.\widehat{r}) (\overrightarrow{S}_{l}.\widehat{r})
-\overrightarrow{S}_{N}.(\overrightarrow{S}_{l}) ]
\label{11}
\end {equation}

Inserting $I_{N}$(nucleon isospin), $I_{H}$ (meson isospin), $S_{N}$ (nucleon spin), $S_{l}$ (the lightest quark's spin in  the meson), and $K=S_{N}+S_{l}$ into the potential, We presented  hadronic molecular structure of two  pentaquarks in  Table 2.\\

In this calculation we have ignored the tensor term $ S_{12} $ for studying the pentaquarks ground state  \cite{24};  hence, we  used the central part of Yukawa potential. Pentaquark binding energy is defined as the energy used when breaking a  pentaquark into its components, i.e. meson and baryon. We adopted the following constants forthe bound state of heavy pentaquarks from [15] (Table 1):\\
\vspace*{0.5cm}
\begin{small}
\begin{center}
Table \ref{1}:  Potential parameters used for the bound state of heavy pentaquarks [15]
\end{center}
\begin{center}
\begin{tabular}{c c}
\hline
Paramerer &Value  \\ 
\hline
\hline 

   $ g _{A}  $              &    1.27             \\ 
$f_{\pi}$  &   131 MeV       \\
$ g _{H}$                &    -0.59             \\
$m_{\pi}$  &   138 MeV       \\ 
$m_{Ni}$  &   938.92 MeV       \\
$m_{B} $ &   5279 MeV       \\
$m_{D}$  &   1867 MeV       \\
\hline
\end{tabular}  
\end{center}
\end{small}
\vspace*{2.5cm}

\begin{small}
\begin{center}
Table 2:Hadronic molecular structure of pentaquarks
\end{center}
\begin{center}
\begin{tabular}{c c}
\hline
Hadronic structure  & \quad  \quad pentaquark (meson+nucleon) \quad   \\ 
\hline
\hline 

$\theta_{c}$  ($uudd\bar{c}$)& \quad \quad   uud /  udd , $d\bar c$  / $u\bar{c} $ \quad   \\ 
$\theta_{b}$ ($uudd\bar{b}$)& \quad  \quad   uud /  udd ,$ d\bar b$  / $u\bar{ b}$ \quad   \\  \hline

\end{tabular}  
\end{center}
\end{small}

\vspace*{1 cm}

 Pentaquark masses are calculated according to equation  (\ref{5}) 
\begin{equation}
M(pentaquark)=m_{meson}+m_{baryon}+E_{b}
  \label{5}
  \end{equation} 
In order to find the binding energy ($E_{b}$), first we solve Lippmann-Schwinger  equation  for the two-body system. In this approach, the kernel is diagonized and the eigenvalue spectrum is identified (spin-spin interaction in the potential and spin splitting are ignored). The data required include reduced mass of  mesons and nucleons, proposed binding energy, potential co-efficients,  $ r_{0}$=1 or 1.5 fm   and  r-cutoff=20 fm. Table 3 shows the binding energies that we found.\\
 \begin{small}
\begin{center}
Table 3: The calculated binding energies (MeV)  for $\theta_{c}$ (D meson) and $\theta_{b}$ (B meson)  for $ I=0 $. Column A: $  V_{0}$=-276 MeV and $r_{0}$=1 fm;   Column B: $  V_{0}$=-62.79 MeV and $ r_{0}$=1.5 fm
\end{center}
\begin{center}
\begin{tabular}{c ccccc}
\hline
& pentaquark& \quad A& \quad B& \quad \quad A[15]  & \quad  \quad B[15]\quad \quad \\ 

\hline
\hline 

&$\theta_{c}$  &\quad 113.32  & \quad  8.21& \quad \quad  113.99 &\quad \quad  8.45  \quad  \quad \\ 

&$\theta_{b}$ & \quad 140  & \quad 15.1 & \quad  \quad 139.48 & \quad \quad  15.46 \quad \quad \\  \hline
\\
\end{tabular}  
\end{center}
\end{small}
We adopted Yukawa potential from [15] and our calculated energies are in good agreement with [15]. We present pentaquark masses in Table 4. \\

 \begin{small}
\begin{center}
Table 4: Pentaquark calculated masses(MeV)  for $ I=0 $.
\end{center}
\begin{center}
\begin{tabular}{c ccccccc ccc c}
\hline
& pentaquark& &&& Mass(A)&&& & Mass(B)&& Mass in[25,26,27]\\ 

\hline
\hline 

&$\theta_{c}$ 
&& & & 2665.92  &  &&&2790.82  &&2650\\

&$\theta_{b}$  &&& & 6104.6  &  & && 6209.71& & 6391\\   \hline
\end{tabular}  
\end{center}
\end{small}

\vspace*{1cm}
\section{Results and Discussion}
In this paper, we solved Lippmann-Schwinger equation for pentaquark systems. We managed to obtain the binding energy and used it to calculate the masses of these systems. The pentaquark is considered as the bound state of a  baryon and a heavy meson. We
used a  baryon-meson picture to reduce a complicated five-body problem to a simple two-body problem. In Table 4,  we have listed our numerical results for masses of pentaquark systems, and pentaquark masses are compared with the results obtained in \cite{25,26,27}.  \\

 Our method is appropriate for investigating  tetraquark systems too. In our previous work,  we  investigated   the  tetraquark  as  the  bound  state  of a  heavy-light  diquark  and  antidiquark.      We  used  the  diquark-antidiquark  picture  to  reduce  a  complicated  four-body  problem 
to a   simple two-body problems.\\
 We analyzed diquark-antidiquark  in  the  framework  of  a  two-body  (pseudo-point) problem. We made use  of  the potential  coefficients proposed  by  Ebert  et  al. We  solved  Lippmann-Schwinger equation numerically for charm diquark-antidiquark systems and found the eigenvalues to calculate the binding energies and masses of heavy tetraquarks with hidden charms \cite{28}.


\vspace*{1cm}

\section*{Acknowledgments } 
This work is supported by University of Kashan under Grant No. 65500.6. We also express our greatest gratitude to Dr. M.R. Hadizadeh for his helpful technical comments on the numerical solution of the LS integral equation and also for comments that helped us to revise the text of the paper.\\

\section*{Conflict of Interests}
The authors declare that there is no conflict of interests regarding the publication of this paper.

\end{document}